\documentclass[]{aastex631}

\newcommand{\hetg}{{\small {\it Chandra}/HETG}}
\newcommand{\resolve}{{\small XRISM/{\it Resolve}}}
\newcommand{\kms}{km\,s$^{-1}$}
\newcommand{\NH}{$N_{\rm H}$}

\newcommand{\gro}{{\small GRO\,J1655-40}}
\newcommand{\grs}{{\small GRS\,1915+105}}
\newcommand{\gx}{{\small GX\,13+1}}
\newcommand{\fu}{{\small 4U\,1630-472}}

\begin{document}

\title{Elemental Abundances in X-ray Binary Outflows}

\author[0000-0003-2535-6436]{Noa Keshet}
\affiliation{Department of Physics, Technion
Haifa 32000, Israel}

\author[0000-0001-9735-4873]{Ehud Behar}
\affiliation{Department of Physics, Technion
Haifa 32000, Israel}

\author[0000-0003-2869-7682]{Jon M. Miller}
\affiliation{Department of Astronomy, University of Michigan, Ann Arbor, MI 48109, USA}



\begin{abstract}
Line resolved X-ray spectra of outflows from X-ray binaries are interesting since they provide quantifiable measures of the accreted material on to the compact object (black hole or neutron star), which can not be observed directly in the accretion disk.
One such measurement that has been largely overlooked is that of the elemental abundances, which potentially provide insights into the origin of the ejected material.
Using the \hetg\ grating spectrometer 
we measure and present elemental abundances in 
four low-mass X-ray binaries.
We compare two measurement methods. 
One is by fitting line series of individual ions and reconstructing the absorption measure distribution (AMD), and the other is a global fit with one or two individual ionization components.
All outflows feature a steep AMD strongly favoring high ionization degrees.
The present abundances are consistent with previous works suggesting the abundances in the outflows are non-solar.
We find a tentative trend of increasing abundances with atomic number, which fits some core-collapse supernova models, but no exact match to a specific one. 

\end{abstract}



\section{Introduction} \label{sec:intro}

Low-mass X-Ray Binaries (LMXBs) are known to be neutron stars (NSs) and stellar black holes (BHs) accreting mass from a low-mass main-sequence companion. 
Outflows in these sources have been studied for more than twenty five years \citep{Ueda1998, Kotani2000}. 
They are observed primarily in the so-called high, soft state, where the disk dominates the X-ray emission.
They are thus generally understood to be lifted off the disk into our line of sight, and observed at a high inclination angle with respect to the disk axis.
The exact physical launching mechanism remains to be determined.
LMXB outflows tend to feature the most ionized species, up to He- and H- like Fe. 
In the present millennium, this made the \hetg\ spectrometer \citep{Canizares2005} the instrument of choice for studying these outflows \citep[some examples of which are][]{Lee2002, Ueda2004, Miller2006, Neilsen2009,  Marshall2013, Gatuzz2019}. 
For more information, see reviews by \citet{DiazTrigo2016, Neilsen2023}. Recently, the \textit{Resolve} spectrometer on board XRISM \citep{Tashiro2020} started providing more exquisite spectra of these outflows (XRISM collaboration 2025, in preparation). 

Many open questions about X-ray binary outflows still remain. 
The temporal variability of these binaries imposes a challenge, making observations of distinct outflows with rich absorption line spectra scarce. 
Consequently, attempts to study a sample of sources on an equal footing, for common phenomenology are lacking. 
Specifically relevant to the present work, the elemental abundances in the winds have been largely assumed to be solar and not investigated in detail.
Abundances in X-ray binary outflows have the potential of shedding light on the origin of the gas, whether from the companion star, or from residuals of the events leading to the formation of the compact object, either a NS or a BH. 
Metal enrichment in the universe is the result of supernova (SN) explosions, which are also believed to be the progenitors of stellar BHs and NSs. The material left behind thus pollutes the binary environment, and its residuals could appear in the outflows observed. 
A known blind spot of X-ray spectra is the lack of H lines. Hence, all abundance measurements are relative, and not absolute. Common practice is to assume one abundance has a solar abundance (e.g., Fe) and to measure all other abundances relative to it.

There are just a few previous works that report elemental abundances in LMXB outflows.  
\citet{Ueda2004} measured abundances of eight different elements - Fe, Mn, Cr, Ca, Ar, S, Si, Mg -  in the NS LMXB \gx\ by fitting their K$\alpha$ absorption lines. They found the abundances to be within a factor of 1.5 of the solar values, while \citet{Allen2018} analyzed the same spectra, using global fits, and found over abundance factors with respect to Fe of $2-6$. 
\citet{Lee2002} analyzed the spectrum of the BH LMXB \grs\ and reported anomalous abundances of Fe, S, Si, Mg based on the their neutral K edges. 
\citet{Marshall2013} found an over-abundance with respect to solar of all metals by a factor six, and an especially high abundance of Ni (factor 15) in a spectrum of the BH LMXB SS\,433, which they suggested originated in the SN explosion preceding the compact object. 
\citet{Kallman2009} analyzed the spectrum of the BH LMXB \gro\ with a global fit, using a single ionization component, to find mostly solar abundances.
However, recently we reported abundances of eighteen different elements in the same outflow, finding an overabundance of odd-Z elements \citep{Keshet2024}.
Interestingly, both of these results differ dramatically from those measured in the atmosphere of the \gro\  companion.

No clear picture emerges from these works on abundances in LMXB outflows, perhaps because the applied methods vary drastically between authors. 
In this work we perform a uniform analysis of outflows in \grs , \gx , and \fu , 
measured with \hetg , and added to the results of \gro .
We utilize their rich spectra, which include well resolved blue-shifted absorption lines from many elements at different levels of ionization. To characterize the ionization level of the wind in steady state we use the ionization parameter defined as

\begin{equation}
    \xi = \frac{L}{nr^2}
\label{eq:xi}
\end{equation}

\noindent where $L$ is the ionizing luminosity, $n$ the H number density, and $r$ the distance from the ionizing source. $\xi$ represents the balance between photo-ionizing flux ($\propto L/r^2 $) and recombination ($\propto n$) rates. For each target the ionization distribution is considered through the different ions resolved in the spectrum, and the abundances are measured while taking into account this distribution. 
These abundances are then compared to those obtained with global fit models using SPEX.
The goals of this paper are to reach universal conclusions about the ionization distribution and elemental abundances in outflows of LMXBs, to learn about individual binaries and perhaps their past, and to understand the discrepancies between abundances obtained by different methods. 

\section{Observations and Data} 

The small sample for the present work is constructed based on archival \hetg\ spectra of LMXB outflows, in which we could identify absorption lines from at least seven different elements.
Another criterion crucial for the present method is the presence of lines from both H-like and He-like ions of the same element. We require this for at least two different elements (see \ref{sec:EWmethod}). 
For all targets, we use the first and third diffraction order to maximize the spectral resolution of {\small HETG}. 
When possible, several spectra that appear to represent the same outflow are analysed simultaneously. This is determined by comparing the absorption lines and by restricting the time between observations to up to two weeks. 
For targets with many outflow observations over several years, such as GRS\,1915+105, we focused on a single outburst event. 
The targets and observations analyzed in this work are detailed in Table\,\ref{tab:obs}. The results for \gro\ are taken from \citep{Keshet2024}.

\begin{deluxetable}{lcccccc}
\tabletypesize{\scriptsize}
\tablewidth{0pt}
\tablecaption{\hetg\ Observations Used in the Present Work
}
\label{tab:obs}
\tablehead{
\colhead{XRB} 
& \colhead{Observations ID} 
& \colhead{Start Date} 
& \colhead{Exposure} 
& \colhead{Total Exposure}
& \colhead{Previous HETG} \\
\colhead{}
& \colhead{}
& \colhead{}
& \colhead{(s)}
& \colhead{(s)}
& \colhead{Absorber Analysis}
}
\startdata
{GRO\,J1655-40} & 5461  & 2005 Apr 1  & 26218    & 26218 & \cite{Miller2006,Miller2008,Tomaru2023, Keshet2024}\\
\hline
{GRS\,1915+105} & 7485  & 2007 Aug 14 & 48759    & 48759 & \cite{Lee2002,Ueda2009,Miller2016} \\
\hline
{}              & 11814 & 2010 Aug 1  & 28119    &  &\\
{}              & 11815 & 2010 Jul 24 & 28119    &  & \\
{GX\,13+1}      & 11816 & 2010 Jul 30 & 28119    & 136896 & \cite{Ueda2004}\\
{}              & 11817 & 2010 Aug 3  & 28120    &  &\cite{Allen2018} \\
{}              & 11818 & 2010 Aug 5  & 24417    &  \\
\hline
{}              & 13714 & 2012 Jan 17 & 28926    & & \cite{Gatuzz2019} \\
{4U\,1630-472}  & 13715 & 2012 Jan 20 & 29283    & 116926 &\cite{Trueba2019}\\
{}              & 13716 & 2012 Jan 26 & 29280    &  \\
{}              & 13717 & 2012 Jan 30 & 29438    &  \\
\enddata
\end{deluxetable}

Table\,\ref{tab:params} lists the physical parameters of the LMXBs from the literature with their references. The table details compact object mass, companion mass and the compact object type (BH or NS). 

\begin{deluxetable}{lcccccc}
\tabletypesize{\scriptsize}
\tablewidth{0pt}
\tablecaption{Source Physical Parameters
}
\label{tab:params}
\tablehead{
\colhead{XRB} 
& \colhead{Distance} 
& \colhead{Compact Object Type} 
& \colhead{Compact Object Mass} 
& \colhead{Companion Mass}\\
\colhead{}
& \colhead{(kpc)}
& \colhead{}
& \colhead{($M_{\odot}$)}
& \colhead{($M_{\odot}$)}
& \colhead{}
}
\startdata
{GRO\,J1655-40} & 3.2  & Black Hole  & $6.6 \pm 0.5$\,\citep{Shahbaz2003} & $2.8 \pm 0.3$\,\citep{Shahbaz2003} \\
{GRS\,1915+105} & 11 & Black Hole  & $10.1 \pm 0.6$\,\citep{Steeghs2013} & $0.5 \pm 0.3$\,\citep{Steeghs2013} \\
{GX\,13+1}      & 7  & Neutron Star & 1.4                              & $1.2$\,\citep{Allen2000}* \\
{4U\,1630-472}  & 4.7-11.5 & Black Hole  & $10.0 \pm 0.1$\citep{Seifina2014} & not measured \\
\enddata
\tablenotetext{*}{Based on K5III identification \citep{Bandyopadhyay1999}}
\end{deluxetable}

\section{Method}\label{sec:method}

In order to obtain elemental abundances we need to first understand the ionization distribution in the outflow, to factor in the fractional abundances of the observed ions, which depend on $\xi$. 
We compare two different approaches for the ionization distribution. 
In the first one, we reconstruct a continuous ionization distribution from individual ionic measurements \citep{Holczer2007, Keshet2024}.
In the second, we fit the spectrum globally with two discrete $\xi$-components, using SPEX \citep{Kaastra2024}.    

\subsection{Ionic Column Densities}\label{sec:EWmethod}

The elemental abundances relative to H ($A_{\rm Z}$) are calculated from the measured absorption line strengths, which represent ionic column densities $N_{\rm ion}$. 
Consider the following general relation 

\begin{equation}
    \label{eq_Nion}
    N_{\rm ion} = A_{\rm Z} \int f_{\rm ion}(\xi) \frac{dN_{\rm H}}{d\log \xi} d\log \xi
\end{equation}

\noindent where $N_{\rm H}$ is the hydrogen column density, $\xi$ is the ionization parameter (Eq.\,\ref{eq:xi}), and $f_{\rm ion}(\xi)$ is the ionic fractional abundance. $dN_{\rm H} / d\log \xi$ is the Absorption Measure Distribution \citep[AMD,][]{Holczer2007}, which represents the broad ionization distribution of column in the outflow.
The $f_{\rm ion}$ values are taken from photoionization balance calculations, computed separately for each individual target based on its broadband ionizing continuum (SED). 
In the present work we use the Cloudy code \citep{Ferland2013} 
for this.
For a fully detailed description of our method, see \citep{Keshet2024}.

The equivalent width $EW$ is the integral over the 
absorption line profile 
$EW = \int (1-e^{-\tau(E)})dE$, where $\tau (E)$ is the optical depth. 
The $EW$ can be measured directly from the spectrum.
Its curve of growth $EW(N_{\rm ion})$ yields $N_{\rm ion}$.
The ionic column densities are measured with the ion-by-ion fitting code \citep{Peretz2018}, which implements the curve of growth method for all lines and simultaneously fits for the $N_{\rm ion}$ values and 
a global velocity width.
The third order grating spectra are also utilized to constrain the line widths, which are parameterized by
$v_{\rm turb} = \sqrt2 \sigma_v$, where $\sigma_v$ is the Gaussian line profile standard deviation. 
For \grs\ $v_{\rm turb}= 140\,$\kms\ is adopted.
For \gx\ $v_{\rm turb}= 420\,$\kms\, and for \fu\ $v_{\rm turb}= 560\,$\kms\ .
Where possible, $N_{\rm ion}$ is measured both with the curve of growth on strong lines, and the ion-by-ion fit, to validate consistency. The final values used in the analysis are those measured by the ion-by-ion code, 
which are listed in Table\,\ref{tab:ions} for all targets in the sample.

\begin{deluxetable}{lccccccc}
\tabletypesize{\scriptsize}
\tablewidth{0pt}
\tablecaption{Ions identified in the HETG spectrum of the targets in the sample and used for abundance estimates. 
}
\label{tab:ions}
\tablehead{
\colhead{Ion} 
& \colhead{} 
& \colhead{$N_{\rm ion}$} 
& \colhead{}\\
\colhead{} 
& \colhead{} 
& \colhead{$10^{17}$[cm$^{-2}$]} 
& \colhead{} \\
\colhead{} 
& \colhead{\grs} 
& \colhead{\gx} 
& \colhead{\fu} 
}
\startdata
{$\rm Na^{+10}$} & $0.23 \pm 0.22$  & -              & -             \\             
{$\rm Mg^{+11}$} & $0.8 \pm 0.2$    & $0.20 \pm 0.07$& -             \\
{$\rm Al^{+12}$} & $0.20 \pm 0.09$  & -              & -             \\
{$\rm Si^{+13}$} & $1.6 \pm 0.2$    & $0.51 \pm 0.05$& $0.8 \pm 0.2$ \\
{$\rm Si^{+12}$} & $0.4 \pm 0.3$    & $0.04 \pm 0.02$& $0.16 \pm 0.10$\\
{$\rm S^{+15}$}  & $1.9 \pm 0.1$    & $0.38 \pm 0.07$& $0.5 \pm 0.1$\\
{$\rm S^{+14}$}  & $0.4 \pm 0.1$    & $0.14 \pm 0.05$& $\leq0.1$      \\
{$\rm Ar^{+17}$} & $1.28 \pm 0.06$  & $0.24 \pm 0.06$& $0.36 \pm 0.07$\\
{$\rm K^{+18}$}  & $0.09 \pm 0.08$  & -              & -             \\
{$\rm Ca^{+19}$} & $1.5 \pm 0.2$    & $0.4 \pm 0.2$  & $0.71 \pm 0.09$\\
{$\rm Ca^{+18}$} & $0.13 \pm 0.05$  & -              & -             \\
{$\rm Ti^{+21}$} & $\leq0.1$        & -              & -             \\
{$\rm Cr^{+23}$} & $1.3 \pm 0.4$    & $0.4 \pm 0.1$  & $0.4 \pm 0.2$ \\
{$\rm Cr^{+22}$} & $0.4 \pm 0.2$    & $\leq 0.3$     & -             \\
{$\rm Mn^{+24}$} & $1.7 \pm 0.5$    & -              & $0.3 \pm 0.2$ \\
{$\rm Mn^{+23}$} & $0.7 \pm 0.5$    & -              & $0.2 \pm 0.1$ \\
{$\rm Fe^{+25}$} & $50 \pm 6$       & $13.3 \pm 0.9$ & $110 \pm 40$  \\
{$\rm Fe^{+24}$} & $22 \pm 4$       & $1.8 \pm 0.2$  & $21 \pm 5$    \\
{$\rm Fe^{+23}$} & $0.5 \pm 0.2$    & -              & - \\
{$\rm Ni^{+27}$} & -                & -              & $3.0 \pm 1.0$\\
{$\rm Ni^{+26}$} & $3.0 \pm 2.0$    & -              & $1.0 \pm 0.5$\\
\enddata

\end{deluxetable}

\begin{figure}
    \centering
    \includegraphics[width=\textwidth]{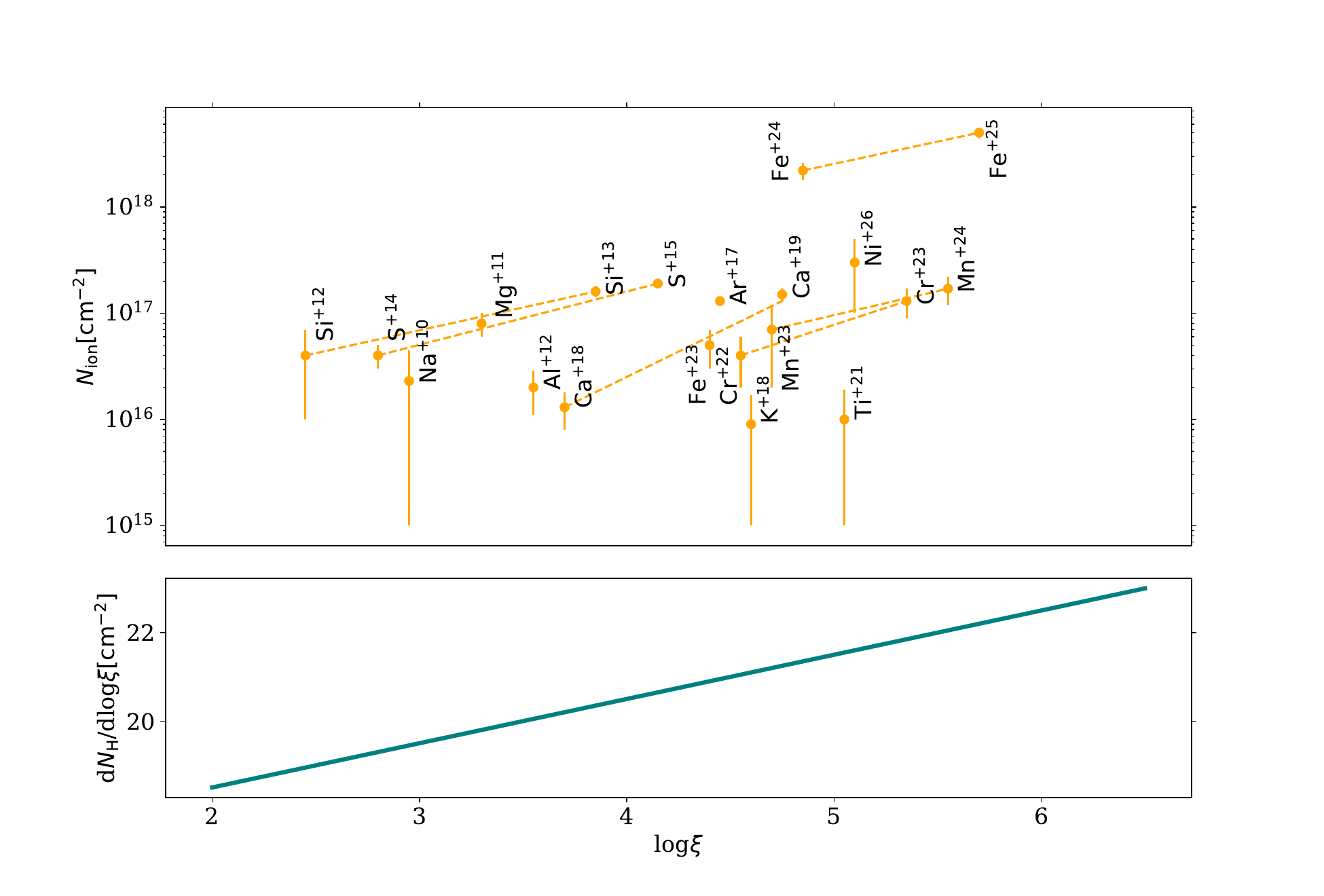}
    \caption{(\textit{top panel}) Distribution of measured ionic column densities $N_{\rm ion}$ in the \grs\ outflow, each plotted at its $\xi$ of maximal formation. 
    This outlines the overall distribution of $N_{\rm H}$ with $\xi$, showing a gradual increase with $\xi$. The dashed lines represent the local slopes from H-like and He-like ion pairs, which are used to reconstruct the continuous AMD in Eq.\,\ref{eq_Nion} (\textit{bottom panel}).  
   Since there is no H in the spectrum, H-like Fe is arbitrarily chosen as the reference point.}
    \label{fig:GRS_AMD}
\end{figure}

To obtain elemental abundances $N_{\rm ion}$ is not enough, we also need the AMD (Eq.\ref{eq_Nion}). 
The AMD shape is unknown a-priori, and needs to be constructed from the the multitude of $N_{\rm ion}$ measurements. We assume the AMD is a simple power-law, which can be broken into smaller segments when needed. 
We require the AMD to satisfy the condition that two ions of the same element yield the same $A_{\rm Z}$ in Eq.\ref{eq_Nion}. 
The more such pairs of ions, the better constrained is the final AMD.
The process of reconstructing the AMD is demonstrated for \grs\ in Fig.\,\ref{fig:GRS_AMD}. In the top panel of Fig.\,\ref{fig:GRS_AMD} we plot all of the measured $N_{\rm ion}$ values used for \grs\, as detailed in Table\,\ref{tab:ions}. The bottom panel is the AMD reconstructed such that it will fulfill the requirement of a consistent $A_{\rm Z}$ for ion pairs of a single element.

\subsection{Global Fitting}

For comparison, we use a second method to obtain abundances, i.e., a global fit with the pion model in SPEX \citep{Kaastra2024}, which has the elemental abundances, $\xi$, $N_{\rm H}$, $v_{\rm out}$ and $v_{\rm turb}$ as free parameters.
Several pion components can be added to obtain an acceptable fit.
The spectral model for each target is comprised of a few continuum components absorbed by pion components.
The continuum is modeled with a black body disk, and a Comptonized component.
The neutral absorption is also accounted for using the "hot" component model. 
We included two free pion absorption components, namely two $\xi$ components that are supposed to represent the AMD, or at least be compared with it. The fit for any additional $\xi$ components preferred zero column density. For \fu\ we weren't even able to constrain a second component. 
For each target, the elemental abundances were tied between the $\xi$ components. 

The high columns and extraordinary signal found in \grs\ requires the inclusion of re-emission from the wind \citep{Miller2016}, so a pion emission component was coupled to each absorption component.   The spectrum of \fu\ required only one pion absorption ($\xi$) component. In order to obtain relative abundances, the Fe abundance is fixed to its solar value, while all other abundances identified in the wind are free to vary in the fit. The best-fit continuum parameters of the global fit models are detailed in Table\,\ref{tab:continuum}. The optical depth for the Comptonization component is fixed at $10^{-3}$ for all targets.

\begin{deluxetable}{lccccccccc}
\tabletypesize{\scriptsize}
\tablewidth{0pt}
\tablecaption{Best-fit continuum parameters for global models 
}
\label{tab:continuum}
\tablehead{
\colhead{}  
& \multicolumn{2}{c}{Statistics}
& \colhead{Galactic Absorption}
& \multicolumn{2}{c}{Comptonization} 
& \multicolumn{2}{c}{Disk black body}\\
\colhead{Target} 
& \colhead{$C_{\rm stat}$}
& \colhead{d.o.f.}
& \colhead{$N_{\rm H} (\times10^{24}/\rm cm ^{2})$}
& \colhead{Norm $(10^{44}\,\rm ph\,s^{-1}\,keV^ {-1})$}
& \colhead{$kT_{\rm plasma}$ (keV)}
& \colhead{Norm $10^{12} \rm cm^{2}$}
& \colhead{$kT $ (keV)}
}
\startdata
{GRS\,1915+105}& 5717.7 & 2416 & $0.0734\pm0.0004$ & $12.4\pm0.1$ & $20.6\pm0.8$ & $2.32\times10^{-7}\pm0.01\times10^{-7}$ & $1.386\pm0.001$ \\
{GX\,13+1} & 11592.9 & 1354 & $0.0449\pm0.0005$ & $1.1\times10^{-7}\pm0.2\times10^{-7}$ & $1.57\pm0.04$ & $2.1\times10^{-8}\pm0.2\times10^{-8}$ & $1.83\pm0.06$ \\
{4U\,1630-472} & $1186$ & $1181$ & $0.1040 \pm 0.0002$& $6.3\times10^{-2}\pm0.8\times10^{-2}$ & $90\pm70$ & $9.1\time10^{-9}\pm0.2\times10^{-9}$ & $2.91\pm0.01$ \\
\enddata

\end{deluxetable}

\section{Results} \label{sec:results}

\subsection{AMD shape}

\begin{figure}
    \centering
    \includegraphics[width=\textwidth]{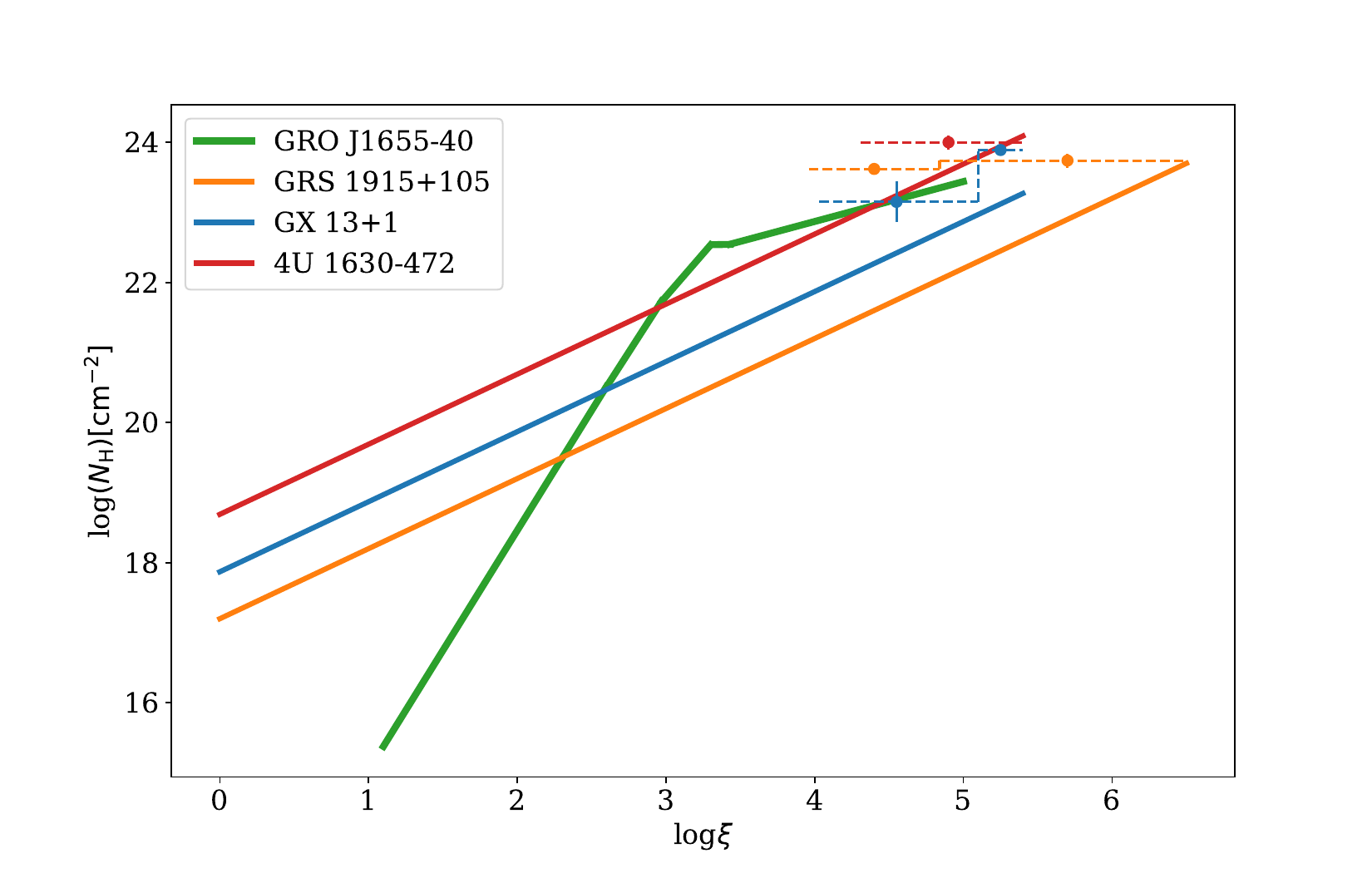}
    \caption{AMDs for all targets in the current sample, including \gro\ from \citet{Keshet2024}. Normalization is determined by assuming a solar  Fe/H abundance for all targets. 
    The dashed lines represent integration over the discrete pion absorption components found in the global fits (Table\,\ref{tab:pion}). The last point for each target represents the total \NH . }
    \label{fig:AMDs}
\end{figure}

We aim for the simplest AMD, namely for the minimal  power-law segments, which still yield consistent  elemental abundances measured from H-like and He-like ions of the same element (Eq.\,\ref{eq_Nion}). 
This is achieved for all three targets using a 
single power-law AMD with a slope of 1.0.
Fig.\,\ref{fig:AMDs} presents the single power-law AMDs reconstructed for the three targets, which are compared with the AMD of \gro\ that required a broken power-law \citep{Keshet2024}. 
For \gx , the single power-law can not reconcile the abundance of S derived from its H-like and He-like ions. Adding a break in the power-law solves this problem, but creates a disagreement between the Si ions. Therefore, we retain the single power-law solution. 
It is curious that the AMDs of the X-ray binary outflows in Fig.\,\ref{fig:AMDs} seem to indicate a universal slope of 1.0.
In the analysis of AMDs in AGN outflows, \citet{Keshet2022} also found a universal slope, but a much shallower one of 0.0-0.5. 
This difference is clearly manifested in the high column densities of low charge states, such as Fe-L and O-K ions, which are typically absent from X-ray binary outflows.

 The results of the global SPEX models are detailed in Table\,\ref{tab:pion}. The discrete ionization components found by these models are also plotted in Fig.\,\ref{fig:AMDs}. Thinking of these components as a distribution, each $N_{\rm H}$ measured in a pion component is added such that the plot ends at the total \NH\ measured.
It can be seen that these one or two ionization components per target do not provide the same sense of increasing column with ionization.
The lack of significant column at low-$\xi$ in the global fits results in their difficulty to measure the low-$Z$ elements (e.g., Na, Mg, Al) that form at lower $\xi$ regions of the outflow.
For \gx , the present global fit is consistent with that of \citet{Trueba2019}, who based on two $\xi$-zones obtained an AMD
$\propto \xi^{0.32\pm0.04}$, between $3.4<\log\xi<5.4$. 

\begin{deluxetable}{lccccccccc}
\tabletypesize{\scriptsize}
\tablewidth{0pt}
\tablecaption{Best-fit absorption parameters for global models using pion components
}
\label{tab:pion}
\tablehead{
\colhead{}   
& \multicolumn{4}{c}{pion 1}
& \multicolumn{4}{c}{pion 2}\\
\colhead{Target}
& \colhead{$N_{\rm H}(10^{24}/\rm cm ^{2})$ }
& \colhead{$\log \xi$}
& \colhead{$v_{\rm out}$ (km/s)}
& \colhead{$v_{\rm turb}$ (km/s)}
& \colhead{$N_{\rm H} (10^{24}/\rm cm ^{2})$}
& \colhead{$\log \xi$}
& \colhead{$v_{\rm out}$ (km/s)}
& \colhead{$v_{\rm turb}$ (km/s)}
}
\startdata
{GRS\,1915+105} & $0.42\pm0.03$ & $3.96\pm0.02$ & $-150\pm 20$ & $100^{+5}_{-5}$ & $0.13\pm0.03$ & $4.84\pm0.06$ & $-930^{-60}_{+150}$ & $150\pm50$ \\
{GX\,13+1}  & $0.14\pm0.04$ & $4.03\pm0.08$ & $-390^{-50}_{+30}$ & $120^{+20}_{-10}$ & $0.63\pm0.04$ & $5.1\pm0.1$ & $-1100^{-100}_{+100}$ &  $120^{+20}_{-20}$\\
{4U\,1630-472}  & $1.0\pm0.1$ & $4.31\pm0.03$ & $-360\pm20$ & $130\pm 5$ & \nodata & \nodata & \nodata & \nodata \\
\enddata
\end{deluxetable}

The total column density $N_{\rm H}$ in all models is obtained by assuming a solar Fe/H abundance. 
A general caveat of measuring $N_{\rm H}$ from absorption lines is the fully ionized plasma at high $\xi$ values that does not produce lines anymore. Some models actually predict a large contribution to $N_{\rm H}$ from this highly ionized gas \citep{Fukumura2017}.
For a meaningful comparison between the methods, we determine a high-$\xi$ AMD cut-off (to keep the AMD finite) where the Fe$^{+25}$ fractional abundance drops to $10\%$ of its peak value. 
This criterion results in $\log\xi$ cut offs of 6.5, 5.4, and 5.4 for \grs , \gx , and \fu , respectively. 
Integrating the AMD up to these $\log\xi$ values yields a total $N_{\rm H}$ for each target of $5\times10^{23}\,\rm cm^{-2}$ for \grs , $1.9\times10^{23}\,\rm cm^{-2}$ for \gx , and $1.2\times10^{24}\,\rm cm^{-2}$ for \fu .
The respective globally fitted values are $5.5\times10^{23}\,\rm cm^{-2}$, $7.7\times10^{23}\,\rm cm^{-2}$, and $1\times10^{24}\,\rm cm^{-2}$. 
Evidently, the agreement for \grs\ and \fu\ is excellent.
The higher column for \gx\ in SPEX arises from modeling the emission of the wind, which is included in the pion models, and not in the ion-by-ion method. The wind geometry in SPEX is predetermined and may not accurately describe the geometry of these outflows.



\subsection{Elemental Abundances}

\begin{figure}
    \centering
    \includegraphics[width=\textwidth]{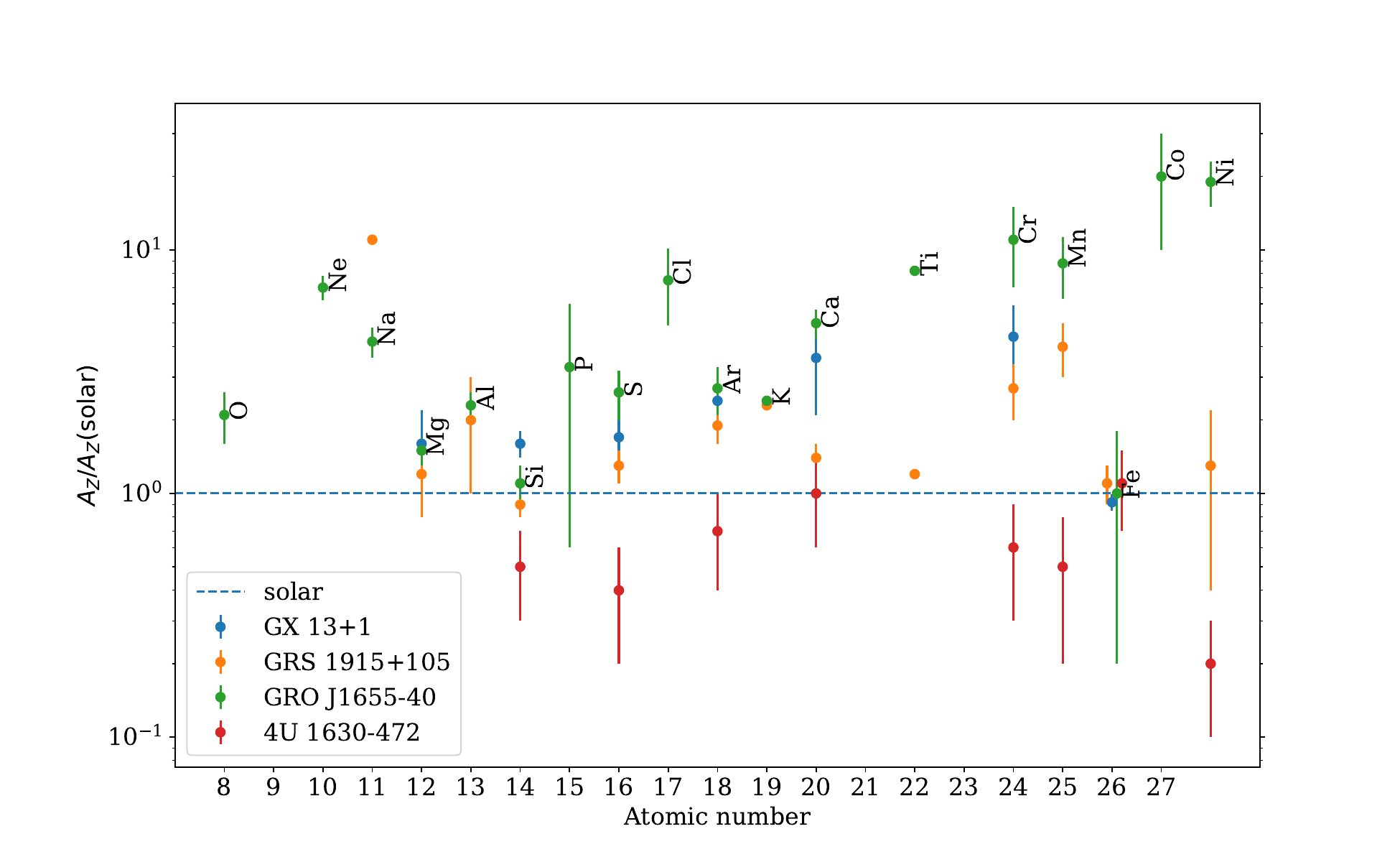}
    \caption{Measured elemental abundances for all targets in the sample, and including \gro\ \citep[from][]{Keshet2024}. Abundances are given relative to Fe and to the solar values. Data points with no error bars represent upper limits. }
    \label{fig:abund}
\end{figure}

Once the AMD is reconstructed, $A_{\rm Z}$ is calculated from each $N_{\rm ion}$, 
using Eq.\,\ref{eq_Nion}. 
The final abundances reported in Table\,\ref{tab:abund} and plotted in Figure\,\ref{fig:abund} are a weighted average of all available ions from each element, with their statistical uncertainties. 
These are compared with the best-fitted abundances of the global models. 
It can be seen that generally, the global fits report smaller uncertainties, as the value of $\xi$ in each component is tightly constrained (see Table\,\ref{tab:pion}), while the ion-by-ion abundances depend on broad, and thus more uncertain AMDs (see Fig.\,\ref{fig:AMDs}).

We consider the possibility that for the strongest lines (of Fe\,K$\alpha$) $N_\mathrm{ion}$ may be higher than what we measure directly, due to the effects of saturation. 
Our analysis does not indicate saturation. However, if troughs are partially refilled by electron scattering, we would be underestimating its effect. Such scattering is also not included in the SPEX global models.
In \gx\, for example, models that included scattering \citep{Tomaru2020} reported a similar total $N_\mathrm{H}$ as those that did not include it \citep{Rogantini2025}. In \gro, \citet{Tomaru2023} include electron scattering and find \NH\ which is an order magnitude higher than all other reports \citep{ Miller2008, Kallman2009, Keshet2024}.  For the three targets studied in this paper, increasing $N_\mathrm{ion}$ for the Fe-K ions by as much as 20\% does not dramatically change the relative abundances beyond their reported uncertainties (Table\,\ref{tab:abund}). 

The abundances of \fu\ show the best agreement between the ion-by-ion method and the global fit. Only Ni whose K-shell lines are at the edge of the \hetg\ band is significantly discrepant.
This is also the outflow with the highest column density and lowest abundances, ranging from sub-solar to solar (w.r.t. Fe).
In \gx , the abundances obtained from the ion-by-ion method are higher, but their uncertainties are relatively large. Both methods suggest an increase in abundance with atomic number reaching super-solar values for Ar, Ca and Cr (w.r.t. Fe).
The results for \grs\ are mixed with some abundances in agreement between methods, and some not. 
The decreasing AMD with $\xi$ implied by the pion components (see Fig.\,\ref{fig:AMDs}) can explain the moderate increase in abundance with atomic number in the global fit, while no obvious such trend is identified in the ion-by-ion measurements, which are based on an increasing AMD with $\xi$.


\begin{deluxetable}{lcccccccc}
\tabletypesize{\scriptsize}
\tablewidth{0pt}
\tablecaption{Elemental abundances for all targets in the current work, based on the ion-by-ion measurements and the global fits. \gro\ abundances are taken from \citet{Keshet2024}. All abundances are presented relative to Fe and given with respect to their solar values.
}
\label{tab:abund}
\tablehead{
\colhead{Element} 
& \colhead{}
& \multicolumn{2}{c}{GRS\,1915+105}
& \multicolumn{2}{c}{GX\,13+1} 
& \multicolumn{2}{c}{4U\,1630-472}
& \colhead{GRO\,J1655-40}\\
\colhead{}
& \colhead{}
& \colhead{Ion-by-ion}
& \colhead{Global fit}
& \colhead{Ion-by-ion}
& \colhead{Global fit}
& \colhead{Ion-by-ion}
& \colhead{Global fit}
& \colhead{Ion-by-ion}
}
\startdata
{O}  & & -            & - & - & -& -  & - & $2.1 \pm 0.5$  \\
{Ne} & &-             & - & - & -& -  & - & $7.0 \pm 0.8$ \\
{Na} & &$11 \pm 10$   & -  & - & -& - & - & $4.2 \pm 0.6$ \\            
{Mg} & &$1.2 \pm 0.4$ &-& $1.6 \pm 0.6$ & - & -& - & $1.5 \pm 0.2$\\
{Al} & &$2 \pm 1$     &- & -   & -& - & - & $2.3 \pm 0.3$ \\
{Si} & &$0.9 \pm 0.1$ & $0.83 \pm 0.07$ & $1.6 \pm 0.2$ & $1.2 \pm 0.1$ & $0.5 \pm 0.2$ & $0.6 \pm 0.1$  & $1.1 \pm 0.2$ \\
{P}  & &-             & - & - & -& -  & - & $3.3 \pm 2.7$\\
{S}  & &$1.3 \pm 0.2$ & $0.82 \pm 0.07$& $1.7 \pm 0.3$ & $1.2 \pm 0.1$ & $0.4 \pm 0.2$ & $0.40 \pm 0.07$ & $2.6 \pm 0.6$ \\
{Cl} & &-             & - & - & -& -  & - & $7.5 \pm 2.6$ \\
{Ar} & &$1.9 \pm 0.3$ & $0.98 \pm 0.08$ & $2.4 \pm 0.7$ & $1.4 \pm 0.1$ & $0.7 \pm 0.3$ & $0.54 \pm 0.08$ & $2.7 \pm 0.6$ \\
{K}  & &$2.3 \pm 2.1$ & $1.6 \pm 0.7$ & -& -& -  & - & $2.4 \pm 2.2$  \\
{Ca} & &$1.4 \pm 0.2$ & $1.01 \pm 0.08$& $3.6 \pm 1.5$ & $1.8 \pm 0.2$ & $1.0 \pm 0.4$ & $0.72 \pm 0.07$ & $5.0 \pm 0.7$  \\
{Ti} & &$1.2 \pm 1.1$ & $1.0 \pm 0.6$ & - & -& - & -  & $8.2 \pm 7.6$ \\
{Cr} & &$2.7 \pm 0.7$ & $1.3 \pm 0.2$ & $4.4 \pm 1.5$ & $1.8 \pm 0.3$      & $0.6 \pm 0.3$ & $0.4 \pm 0.1$ & $11 \pm 4$    \\
{Mn} & &$4 \pm 1$     & $1.0 \pm 0.1$ & -  & -  & $0.5 \pm 0.3$ & $0.1 \pm 0.1$ & $9 \pm 2$     \\
{Fe} & &$1.1 \pm 0.2$ & $1.0$ &$0.93 \pm 0.07$ & $1.0$ & $1.1 \pm 0.4$ & $1.0$ & $1.0 \pm 0.8$\\
{Co} & &-             & - & - & -& -  & - & $20 \pm 10$   \\
{Ni} & &$1.3 \pm 0.9$ & $1.5 \pm 0.3$ & - & -  & $0.2 \pm 0.1$ & $1.2 \pm 0.2$ & $20 \pm 4$    \\
\enddata
\end{deluxetable}

The process of minimization of the C-statistic gives greater weight to the continuum than to any individual line, thus the global fit is less sensitive to specific features. The use of the pion models for absorption favors a minimal number of different $\log\xi$ components, making it insensitive to unique AMD shapes. The use of a minimal number of pion components sample the AMD shape for a small range of $\log\xi$ values and for a limited number of points, making it a crude estimation. Our method demands an AMD that will recreate the same abundances measurements for different ions of the same element, providing self-consistency. This is crucial especially for very rich spectra, as is evident from the case of \gro . There has already been several published attempts at a global fit of this spectrum \citep{Miller2008,Kallman2009,Tomaru2023} with statistically unsatisfying results. Therefore a global-fit for \gro\ was not attempted as part of this work. 

When comparing the shape of the AMDs reconstructed based on the ionic measurements to a simple power-law connecting the pion components, there is no conclusive conclusion. For \fu\ , since there is a single component in the global fit, there is no comparison. The trend of \gx\ is very similar between both methods, with a steep rise towards high $\xi$, and for \grs\ the trend in the global fit is a negative slope. This comparison is problematic due to the low number of points provided by the global fit, therefore the constraints on the AMD shape are loose.


\subsection{Supernova Models}

\begin{figure}
    \centering
    \includegraphics[width=0.49\textwidth]{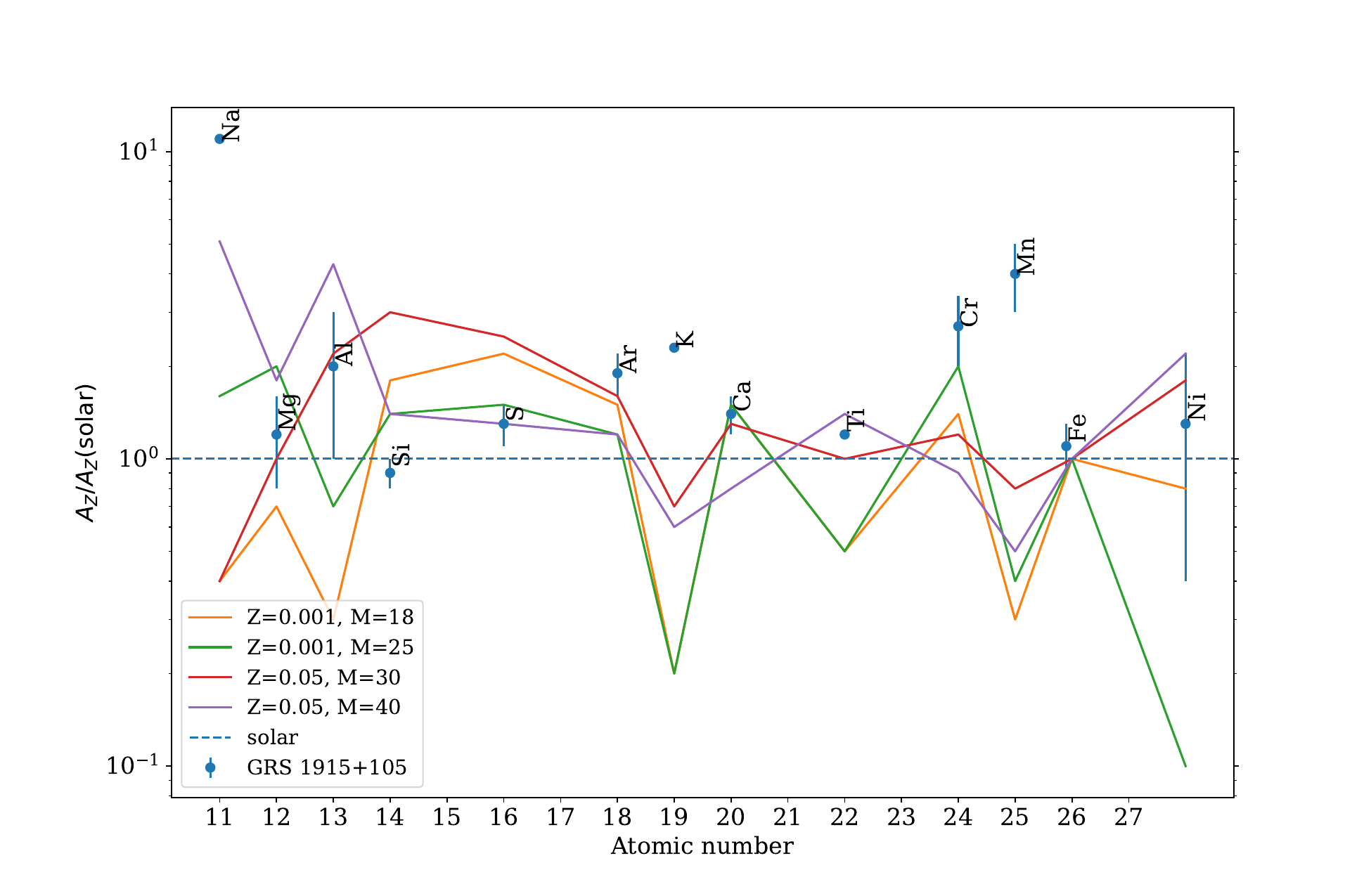}
    \includegraphics[width=0.49\textwidth]{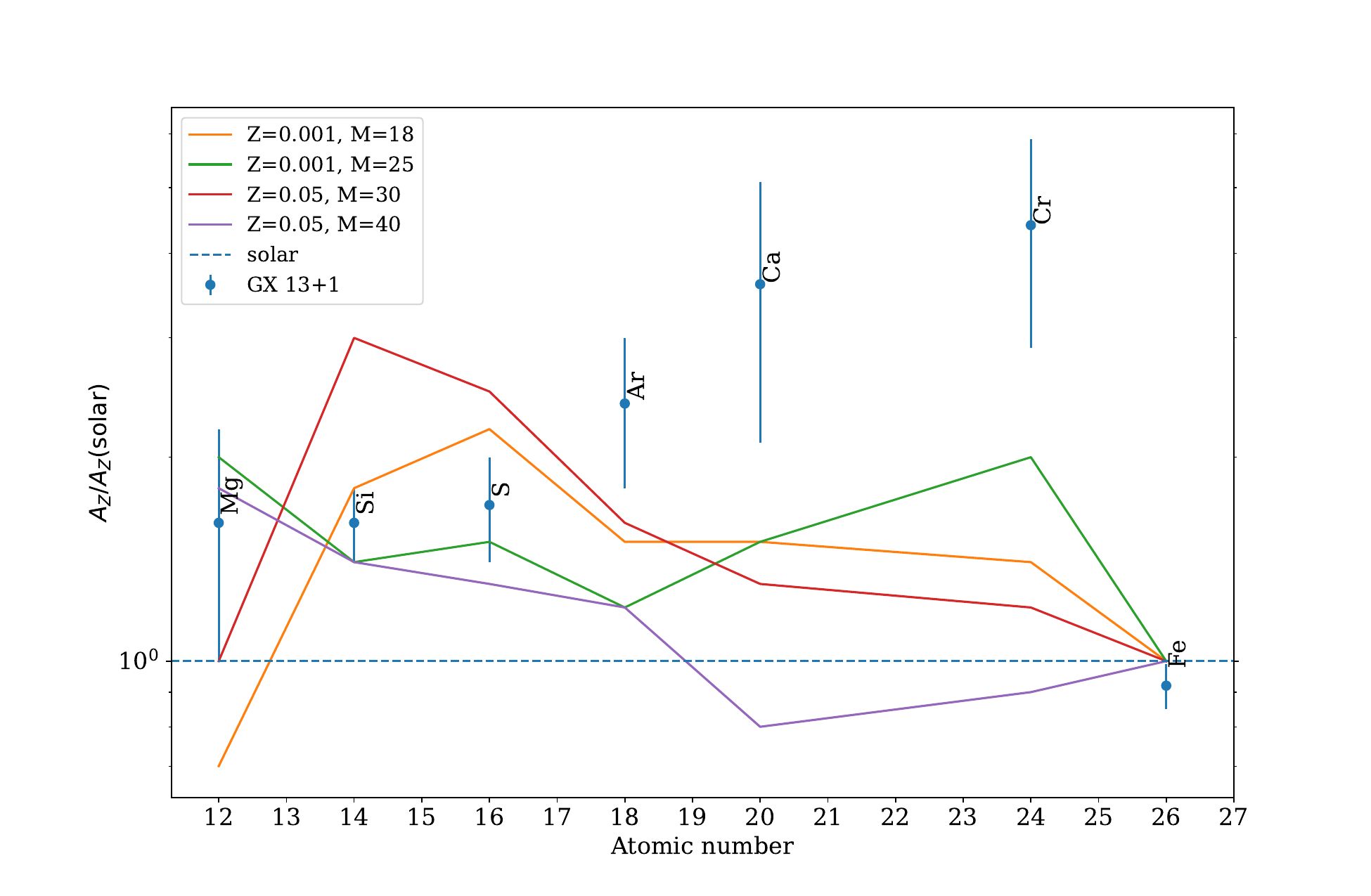}
    \includegraphics[width=0.49\textwidth]{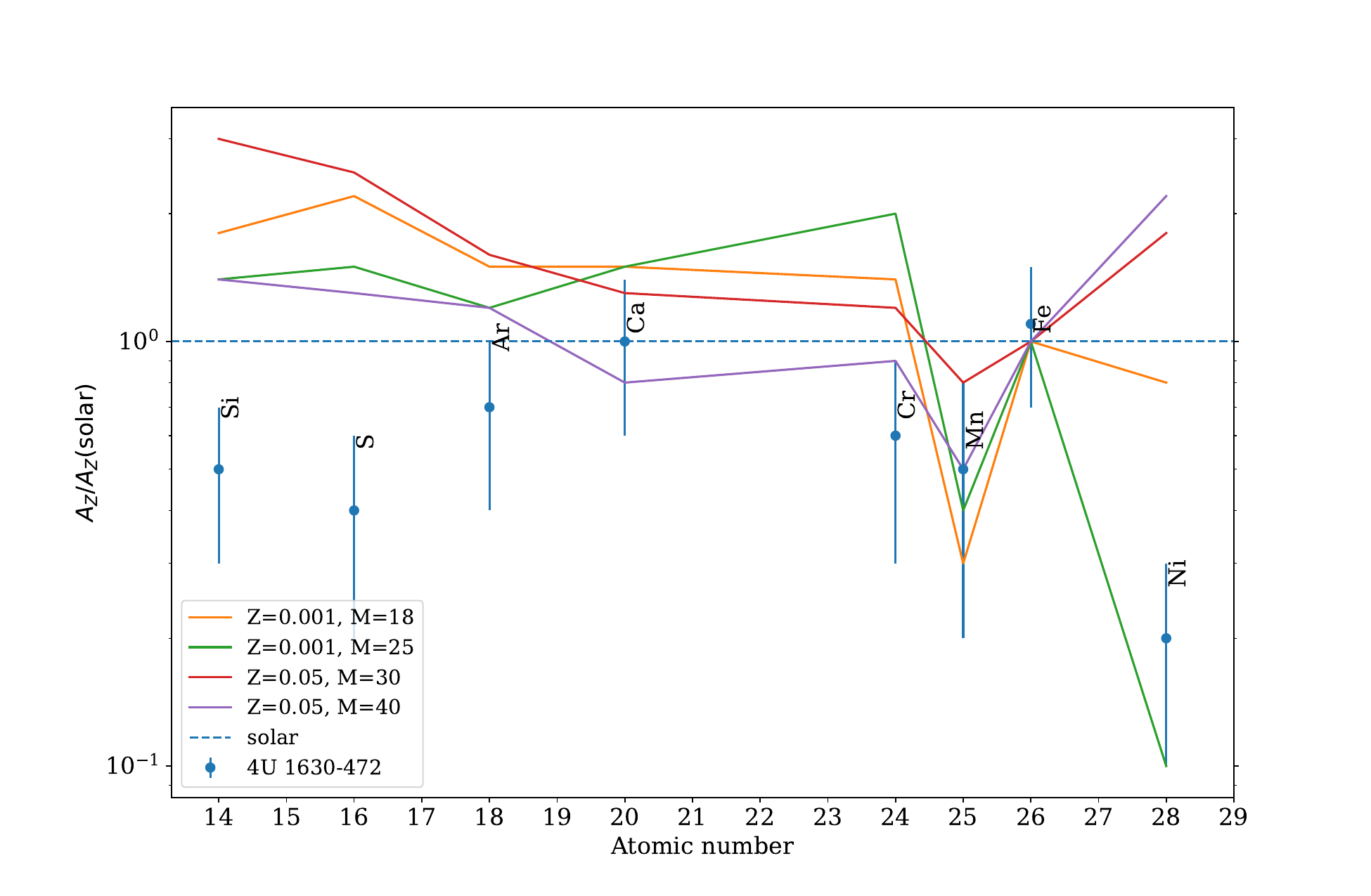}
    \caption{Comparison of the outflow abundances measured with the ion-by-ion method to different SN model yields, taken from \citet{Nomoto2006, Nomoto2013}. The models differ by the initial progenitor mass $M$ and metallicity $Z$. Models are the same in all panels. Lines are drawn between measured elements for each source, just to guide the eye.}
    \label{fig:SNmodels}
\end{figure}

The compact objects in the binaries studied here are understood to have had a massive stellar progenitor, whose core collapsed in a supernova (SN), leaving the NS or BH behind. 
Since these all have low-mass companions, the mass budget of the original binary system is heavily dominated by the progenitor.
To that end, even if a small fraction of the core-collapsed star pollutes the companion, or remains bound to the system, one could expect to see a chemical trace of the SN explosion in the X-ray binary, and its outflow.
Indeed, there is circumstantial evidence for extended Compton thick gas (a photosphere) around X-ray binaries in outburst \citep[e.g.,][XRISM collaboration 2025, in preparation]{Neilsen2016, Shidatsu2016}, which could be SN debris heated by the X-ray source \citep{Keshet2024}.
We surveyed a large range of progenitor masses $M$ and zero-age main-sequence metallicities $Z$ (not to be confused with the atomic number) from \citet{Nomoto2006,Nomoto2013}. 
The elemental abundances measured by the ion-by-ion method are compared in Fig.\,\ref{fig:SNmodels} with SN model yields that best match the present measurements. 
$Z=0.02$ is the solar metallicity, therefore a progenitor with $Z=0.05$ is considered super-solar, and a progenitor with $Z=0.001$ is sub-solar.

None of the SN models fits perfectly any of the outflow abundances. 
The \grs\ abundance pattern resembles that of a high-$M (30\,M_\odot$ or $40\,M_\odot)$ high-$Z (0.05)$ progenitor, except for its high Cr and Mn abundances. In the globally fitted models these are closer to solar (Table\,\ref{tab:abund}), which would improve the agreement.
Conversely, the increase of abundance ratio to Fe with atomic number observed in \gx\ hints towards the yields of a low-$Z (0.001)$ low-$M (18\,M_{\odot}$ or $25\,M_{\odot})$ model. 
The global fits, which show a similar trend, but with slightly lower abundance values would quantitatively fit this SN model better. 
A higher progenitor mass is also ruled out by virtue of \gx\ being a NS binary, and not a BH binary. 
The sub-solar abundance ratios to Fe in \fu\ are atypical of any core collapse SN, hence none of the models agree quantitatively with the abundance values. 
Yet the trend with atomic number is similar to that of \gx , again resembling the $Z=0.001$, $25\,M_{\odot}$ model.
Overall, the far from perfect agreement between the measured values and any of the SN model yields does not enable a conclusion regarding the alleged connection between the material in the outflow and a progenitor.

\section{Discussion and Conclusions} \label{sec:discussion}

The present method of reconstructing an outflow AMD from the ion-by-ion fitting reveals some discrepancies with the more commonly used global fitting methods, in this case with SPEX. 
While the global fits have the obvious advantage of invoking a self-consistent physical state composed of several ionization components, it might not always properly reflect the range of ionization, which is likely present in the outflow.
Conversely, the present AMD realization of a power-law behavior has its own limitations, since there is no reason to assume this shape a-priori.
We find that the abundance trends are recovered by both methods, for example the sub-solar abundances (w.r.t Fe) in \fu\ or the increasing abundance with atomic number in \gx , while some individual abundances, e.g., Cr in \grs\ and in \gx , are significantly discrepant between the methods.
Clearly, a global fit is not sensitive to a specific abundance of a rare element.
If more than one ion is detected from this element, the AMD shape might be a better way to obtain its abundance.

We can compare the presently measured abundances with the few published values in the literature. 
We are not aware of abundance measurements in \fu .
For \gx , \citet[][their Table\,4]{Ueda2004} and \citet[][their Table\,5]{Allen2018}, reported abundances. 
\citet{Allen2018} report over-abundances of a factor of a few for all measured elements compared to Fe, but they do not observe the present increasing trend with atomic number. Their S, Si and Mg abundances are thus higher than those in the present work (Table\,\ref{tab:abund}). 
For \grs , \citet{Lee2002} measured abundances based on neutral edges. They found non-solar abundances of Fe and Si, explaining they could be depleted onto dust, while Mg and S remain at their solar value.
These abundances are in tension with the present finding of approximately solar $A_Z$/Fe ratios for all these elements in the \grs\ outflow (Table\,\ref{tab:abund}).
Indeed, it is possible that the neutral edges are produced at a different location than the ionized outflow. 
In summary, the present results confirm previous measurements of super-solar $A_Z$/Fe abundance ratios in the outflow of \gx , point to sub-solar $A_Z$/Fe abundances in \fu , and more complicated anomalous abundances in \grs . These add to previous non-solar abundances reported from grating spectra in the outflows of SS\,433 \citep{Marshall2013} and \gro\ \citep{Keshet2024}.
Most generally, super-solar abundances are expected from the remnants of core-collapse SN, the massive progenitors of the nowadays observed X-ray binaries. The non-solar abundances generally found in X-ray binary outflows indicate that with improved outflow and SN models, we might be able to learn more about the history of these systems from their abundances.

As new observations of X-ray binary outflows are being carried out with the \textit{Resolve} spectrometer on XRISM, we expect to obtain even better constraints on the high-$\xi$ region of the AMD. The goal will be to repeat this type of analysis for additional outflows with the improved resolution and sensitivity of \resolve .

\begin{acknowledgments}
NK acknowledges the support of a Ramon scholarship from the Israeli Ministry of Science and Technology. EB acknowledges support from NASA grants 80NSSC20K0733, 80NSSC24K1148, and 80NSSC24K1774.
\end{acknowledgments}

%

\vspace{5mm}
\facilities{}


\software{Xspec \citep{Arnaud1996}  
          Cloudy \citep{Ferland2013}      
          }




\bibliography{sample631}{}
\bibliographystyle{aasjournal}



\end{document}